\documentclass[%
 aps,%
 prl,%
 amsmath,amssymb,%
 reprint,%
 superscriptaddress,%
]{revtex4-1}

\newcommand{\NV}{\mathrm{NV}}

\usepackage{graphicx}

\begin{document}


\title{Perfect alignment and preferential orientation of nitrogen-vacancy centers during CVD growth of diamond on (111) surfaces.}

\author{Julia Michl}
 \thanks{main contributor ODMR characterization}
 \affiliation{3. Physikalisches Institut, University of Stuttgart, Germany, Research Center Scope and IQST}
\author{Tokuyuki Teraji}
 \thanks{main contributor CVD growth}
 \affiliation{Environment and Energy Materials Division, National Institute for Materials Science, Tsukuba, 305-0044 Japan}
\author{Sebastian Zaiser}
 \affiliation{3. Physikalisches Institut, University of Stuttgart, Germany, Research Center Scope and IQST}
\author{Ingmar Jakobi}
 \affiliation{3. Physikalisches Institut, University of Stuttgart, Germany, Research Center Scope and IQST}
\author{Gerald Waldherr}
 \affiliation{3. Physikalisches Institut, University of Stuttgart, Germany, Research Center Scope and IQST}
\author{Florian Dolde}
 \affiliation{3. Physikalisches Institut, University of Stuttgart, Germany, Research Center Scope and IQST}
\author{Philipp Neumann}
 \email{p.neumann@physik.uni-stuttgart.de}
 \affiliation{3. Physikalisches Institut, University of Stuttgart, Germany, Research Center Scope and IQST}
\author{Marcus W. Doherty}
 \affiliation{Laser Physics Centre, Research School of Physics and Engineering, Australian National University, Australian Capital Territory 0200, Australia}
\author{Neil B. Manson}
 \affiliation{Laser Physics Centre, Research School of Physics and Engineering, Australian National University, Australian Capital Territory 0200, Australia}
\author{Junichi Isoya}%
 \affiliation{Research Center for Knowledge Communities, University of Tsukuba, Tsukuba, 305-8550 Japan}%
\author{J\"org Wrachtrup}
 \affiliation{3. Physikalisches Institut, University of Stuttgart, Germany, Research Center Scope and IQST}

\date{\today}

\begin{abstract}
Synthetic diamond production is key to the development of quantum metrology and quantum information applications of diamond.
The major quantum sensor and qubit candidate in diamond is the nitrogen-vacancy (NV) color center.
This lattice defect comes in four different crystallographic orientations leading to an intrinsic inhomogeneity among NV centers that is undesirable in some applications.
Here, we report
a microwave plasma-assisted chemical vapor decomposition (MPCVD) diamond growth technique on (111)-oriented substrates that yields perfect alignment ($94\pm2\,\%$) of as-grown NV centers along a single crystallographic direction.
In addition,
clear evidence is found that the majority ($74\pm4\,\%$) of the aligned NV centers were formed by the nitrogen being first included in the (111) growth surface and then followed by the formation of a neighboring vacancy on top.
The achieved homogeneity of the grown NV centers will tremendously benefit quantum information and metrology applications. 
%
\end{abstract}

\maketitle

\section{Introduction}
In recent years many single quantum systems have been explored as qubits and quantum sensors\cite{ladd_quantum_2010,appel_mesoscopic_2009,gierling_cold-atom_2011}.
However, when it comes to realizing quantum technologies several, ideally identical qubits, are needed to build quantum registers and quantum sensors.
This requirement often limits the scaling of quantum devices and the performance of sensing devices.
A well studied single quantum system is the nitrogen-vacancy (NV) defect in diamond, where the qubit or sensor comprises electron and nuclear spins associated with the defect\cite{jelezko_observation_2004-1,jelezko_observation_2004}.
Major advantages of this system are the optical accessibility of single spin quantum states, long spin coherence lifetimes at room temperature and its nanoscopic size\cite{gruber_scanning_1997}.
Several quantum technological benchmark experiments have already been realized with this system, ranging from entanglement generation and storage\cite{dutt_quantum_2007,neumann_multipartite_2008,dolde_room-temperature_2013}, simple quantum computing algorithms\cite{shi_room-temperature_2010} and nanoscale precision measurements of electric\cite{dolde_electric-field_2011} and magnetic fields\cite{balasubramanian_nanoscale_2008,maze_nanoscale_2008} and temperature\cite{toyli_fluorescence_2013-2,kucsko_nanometre-scale_2013,neumann_high-precision_2013}.
Apart from using individually addressable NV centers, these defects can also be used as an ensemble for high precision magnetic field sensors\cite{acosta_diamonds_2009,steinert_magnetic_2013} or as frequency standards\cite{hodges_timekeeping_2013}.
In addition, ensembles of NV centers have been used in hybrid devices where the spins are coupled to the mode of a superconducting microwave resonator\cite{amsuss_cavity_2011,kubo_strong_2010}.
\\
\indent
NV defects in the diamond lattice consist of a substitutional nitrogen atom and adjacent carbon vacancy pair orientated along $\left\langle 111 \right\rangle$ (see fig.~\ref{fig:orientation}a).
Hence, the defects exhibit $C_{3v}$ symmetry and can be aligned along one out of the four $\left\langle 111 \right\rangle$ crystallographic axes.
For purely magnetic interactions between the NV spin and external fields or other spins, the precise order of the nitrogen (N) atom and vacancy (V) along the crystallographic axes does not matter.
However, the four possible crystallographic alignments lead to an inhomogeneity among NV centers and problems in applications where the alignment homogeneity of the NV centers is crucial.
Such applications include those where alignment to an external magnetic field is vital or a NV quantum spin register, where a single misaligned center may render the whole register unusable\cite{neumann_single-shot_2010,dolde_electric-field_2011}.
Indeed, this requirement makes the creation of NV registers extremely demanding.
In a spin ensemble up to 75\% of all spins might be unusable and, worse still, the unusable spins may deteriorate the device performance.
Consequently, a deliberate alignment of NV centers is of great importance for many applications.
Due to refined insight into the NV centers properties \cite{doherty_theory_2012,dolde_detection_2013} the orientation of the N-V pair can now also be determined using the combined application of electric and magnetic fields.
Different orientations for a given alignment cause only minor variations of the spin properties.
\\
\indent
Non-uniform incorporation of impurities among different growth sectors is known in synthetic high pressure and high temperature (HPHT) diamond crystals as well as in CVD diamond 
\cite{dhaenens-johansson_optical_2011,collins_polarised_1989,iakoubovskii_alignment_2004,edmonds_production_2012,ushizawa_boron_1998}.
Thus, homoepitaxial CVD growth using different orientations of HPHT substrates such as (100), (111), (110), (113) is potentially useful for controlling the incorporation of impurities.
Preferential alignment of NV centers along two out of four symmetry-related sites was achieved by using (110)-oriented substrates\cite{edmonds_production_2012} or risers of \{110\} facets in the case of step-flow growth using (100) substrates\cite{pham_enhanced_2012}.
Homoepitaxial CVD growth on (111)-oriented substrates is a potential method to obtain a perfect alignment of color centers having $C_{3v}$ symmetry, such as NV, and $D_{3d}$ symmetry, such as SiV.
While homoepitaxial (111) CVD growth is used for incorporation of phosphorus donors, the growth of high-purity, high-quality (111)-oriented CVD films has not been studied until recently \cite{tallaire_high_2014}.
The crystalline quality of CVD films is affected not only by the densities of dislocations and stacking faults of the HPHT substrates but also by the quality of the polished surface.
High-quality, large HPHT synthetic crystals, which are grown with the (100) growth sectors dominant, are not fitted to obtain (111)-oriented substrates efficiently by cleavage, moreover, the (111) surface is difficult to polish.
\\
\indent
Here, we experimentally demonstrate \textit{perfect alignment} of as grown NV centers along a single crystallographic axis.
Furthermore, we can also show \textit{preferential orientation} along the latter axis.
The NV centers were created by microwave plasma-assisted chemical vapor deposition (MPCVD) diamond growth on a diamond substrate with a (111) surface.
We analyze the CVD diamond layer and substrate by secondary ion mass spectrometry (SIMS) and confocal microscopy.
Optically detected magnetic resonance (ODMR) techniques are applied to the NV centers to verify their \textit{alignment} and \textit{orientation}.
More specifically, ODMR was combined with an axially aligned magnetic field to verify the alignment of the NV centers and ODMR was combined with both transverse magnetic and electric fields to determine the orientations of the NV centers\cite{dolde_detection_2013,doherty_theory_2012}.

\section{Results}
\subsection{\label{sec:CVDgrowth}CVD growth}
\begin{figure}[t]
  \includegraphics[width=1.0\columnwidth]{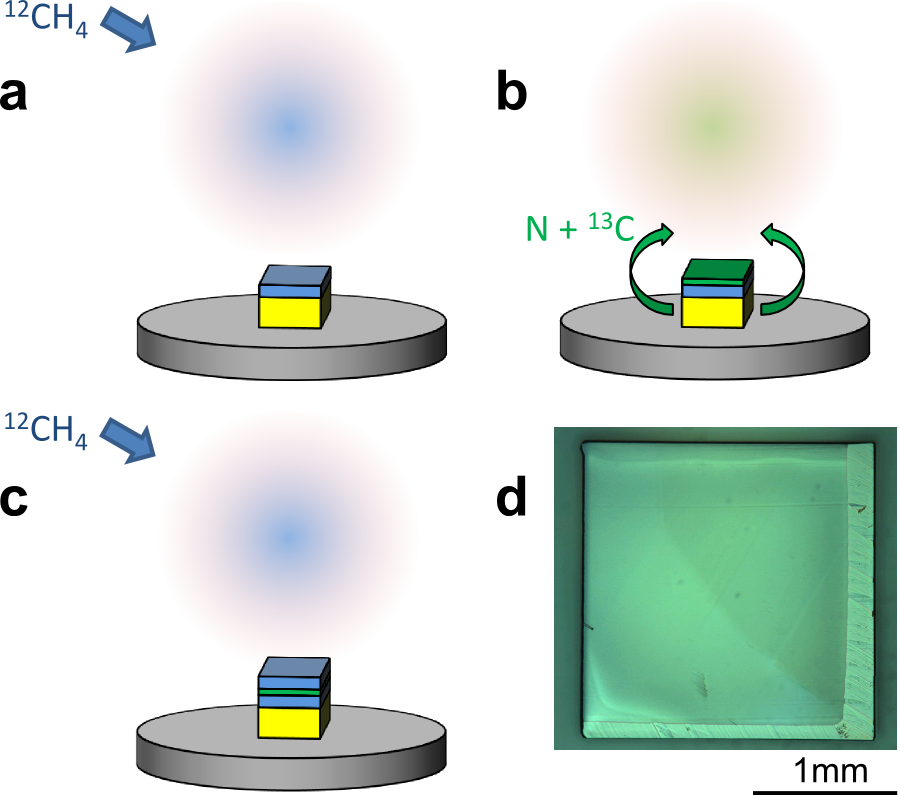}
  \caption{\label{fig:growth}
    \textbf{MPCVD diamond growth.}
		Schematic of growth steps of (111) oriented homoepitaxial diamond growth by MPCVD.
		\textbf{a,} Normal growth mode with $^{12}$C isotopic enriched methane flow.
		\textbf{b,} Etching mode with oxygen flow.
		Nitrogen and $^{13}$C ($1.1\,$\% of total carbon) were etched out from the side wall of the substrate that was exposed to the plasma.
		\textbf{c,} Normal growth mode as in \textbf{a}.
		\textbf{d,} Nomarski-mode optical microscope image of homoepitaxial diamond (111) film grown by MPCVD.
		Substrate had off cut in the direction from the lower right side to the upper left side of this image.
  }
\end{figure}

High-purity homoepitaxial diamond (111) film was deposited using a MPCVD
apparatus (for details see Ref.~\onlinecite{teraji_chemical_2012}). 
A type Ib (111) diamond crystal prepared by HPHT method was used as a substrate (size: $2.0\times 2.0\times 0.5\,$mm$^3$).
It typically contained $100\,$ppm nitrogen.
As source gases we used 9N-grade high-purity hydrogen and $^{12}$C enriched methane with isotopic ratio $99.999\,$\%.
The $^{12}$CH$_4$ contained nitrogen with a concentration of $\sim 3\,$ppm; it was therefore purified using a zirconium purifier.
The residual nitrogen to carbon amount ratio is less than $10\,$ppb (8N-grade).
The total gas pressure, microwave power, methane concentration ratio ($^{12}$CH$_4$/H$_2$), growth duration and substrate temperature employed were $140\,$Torr, $1.2\,$kW, $1\,$\%, $135\,$min, and $1030\pm20\,^{\circ}$C, respectively.
\\
\indent
Since the incorporation efficiency of (111)-oriented crystals is relatively higher than that of (100) crystals, tiny amount of nitrogen need to be introduced into the reactor for the formation of nitrogen related defects. 
In this study, nitrogen in the substrate was intentionally utilized together with unintentionally introduced nitrogen.
During the diamond MPCVD process, CH$_4$ flow was stopped and instead O$_2$ flowed with the concentration ratio (O$_2$/H$_2$) of $2\,$\% for $5\,$min.
Schematic of growth step is shown in fig.~\ref{fig:growth}a-c.
This O$_2$ flow process was repeated 9 times.
\\
\indent
Figure~\ref{fig:growth}d shows optical microscope image of the homoepitaxial diamond (111) film grown by MPCVD.
The surface is rather flat without any non-epitaxial crystallites.
The calculated average roughness was evaluated to be $<\!15\,$nm using a laser microscope equipped with ultraviolet laser.
From the SIMS
measurement, the homoepitaxial layer thickness was estimated to be $7.4\,\mu$m and the $^{12}$C isotopic ratio during the $^{12}$CH$_4$ flow was proved to be $99.99839\pm 0.00002\,$\%, which is the highest among reported carbon isotopic enriched diamond crystals\cite{teraji_effective_2013}.
The growth rate in this condition was $3.3\,\mu$m$\,$h$^{-1}$.
\\
\indent
It was also found that the $^{12}$C isotopic ratio decreased down to $99.99802\,$\% at the depth position where the O$_2$ flow process was applied.
This means $^{13}$C came from the substrate and diamond was grown during the O$_2$ flow mode. 
This phenomenon also suggests that nitrogen is possibly introduced into the plasma, together with carbon, through plasma etching of the Ib substrate.
Please note that the growth during the O$_2$ flow mode was not confirmed for (100) orientation growth.

\subsection{Optical and magnetic resonance analysis}

\subsubsection{Negatively charged NV centers}
The following optical and spin resonance analysis is constrained to the negative charge state of the NV defect (NV$^-$) which is also the desired one in terms of quantum applications.
The electronic ground and first excited state of NV$^-$ is a spin triplet.
Optical illumination leads to subsequent, electron spin state dependent fluorescence and spin polarization\cite{gruber_scanning_1997,aslam_photo-induced_2013}.
This enables single spin ODMR \cite{gruber_scanning_1997}.
\\
\indent
The defect coordinate system is oriented such that the $z$-axis is aligned along the NV symmetry axis, i.e. along one of the crystallographic $\left \langle 111 \right \rangle$ directions.
In addition the $x$-direction lies in one of the center's reflection planes, which contains N, V and a carbon atom adjacent to V (see fig.~\ref{fig:orientation}a).
The spin Hamiltonian relevant for the present work is
\begin{eqnarray}
H &=& D\hat{S}_z^2 + \gamma_{\NV} \underline{B} \cdot \underline{\hat{S}} - \nonumber \\
  & & k_{\perp} \left[ E_x \left( \hat{S}_x^{2}- \hat{S}_y^{2} \right) - E_y \left( \hat{S}_x\hat{S}_{y} + \hat{S}_y\hat{S}_{x} \right)  \right],
\label{eq:H}
\end{eqnarray}
where $\hat{S}_i$ are the $S=1$ spin operators, $D$ is the zero-field splitting of the $m_S=0$ and $\pm 1$ spin sublevels,
$\underline{B}$ is the magnetic field vector,
$\gamma_{\NV}$ is the gyromagnetic ratio of the NV electron spin, and the last term describes the interaction with an external electric field ($k_{\perp} = 170\,\mathrm{kHz\,\mu m\,V^{-1}}$) with transverse components $E_x,\,E_y$ \cite{doherty_theory_2012}. 

\subsubsection{Optical analysis}
\begin{figure}[t]
  \includegraphics[width=1.0\columnwidth]{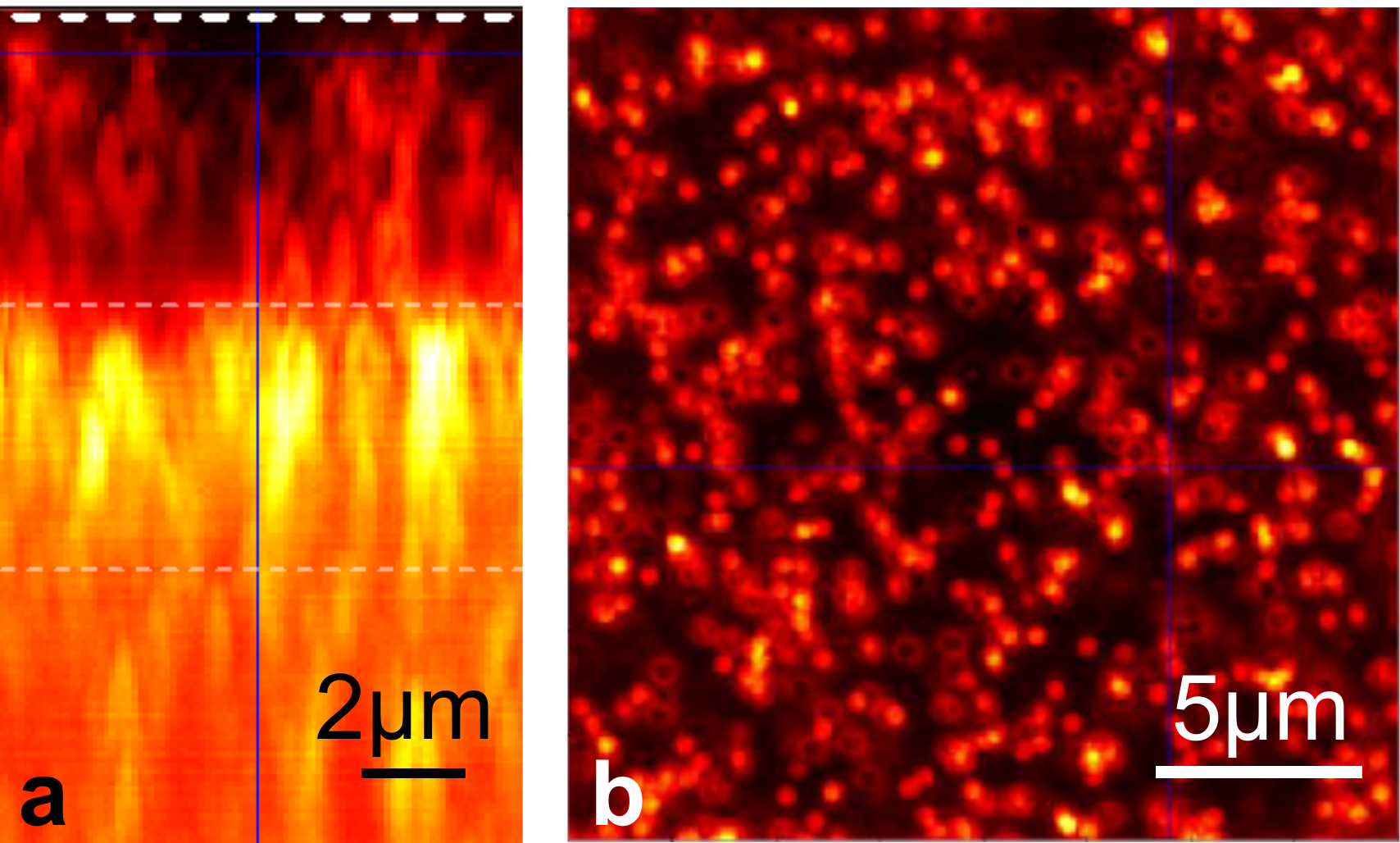}
  \caption{\label{fig:confocal}
    \textbf{Confocal microscopy reveals CVD layer.}
		\textbf{a,} Axial confocal microscopy scan.
		The sample surface is marked by a white dashed line.
		From top to bottom we see the CVD layer and the substrate (see text).
		In vertical direction the image appears shrunk by roughly a factor of two due to the refractive index mismatch at the oil/diamond interface.
		\textbf{b,} Lateral confocal microscopy scan within the CVD layer reveals individually resolvable NV centers.
  }
\end{figure}

The unprocessed, as grown diamond sample is first investigated with respect to the presence of NV centers using confocal laser scanning fluorescence microscopy\cite{gruber_scanning_1997}.
A confocal scan of a plane perpendicular to the diamond surface is shown in fig.~\ref{fig:confocal}a. 
In the top part 
the confocal volume was scanned through the immersion oil.
The interface between oil and diamond (dashed white line) is followed by the less fluorescent CVD layer of $\approx 10\,\mu$m thickness (compare to more accurate SIMS result of $7.4\,\mu$m).
In the CVD layer several fluorescent stripes are visible, which we assign to single NV centers.
Below the CVD layer the diamond substrate is visible exhibiting a much higher density of NV centers.
\\
\indent
In fig.~\ref{fig:confocal}b a lateral confocal scan through the CVD layer reveals many single NV centers.
For the following analysis of NV center alignment and orientation we randomly chose a high number of NV centers within the displayed three dimensional region of the CVD layer.
We can not exclude that few of the chosen NV centers lie already within the substrate.

\subsubsection{Magnetic resonance}
\begin{figure}[t]
  \includegraphics[width=1.0\columnwidth]{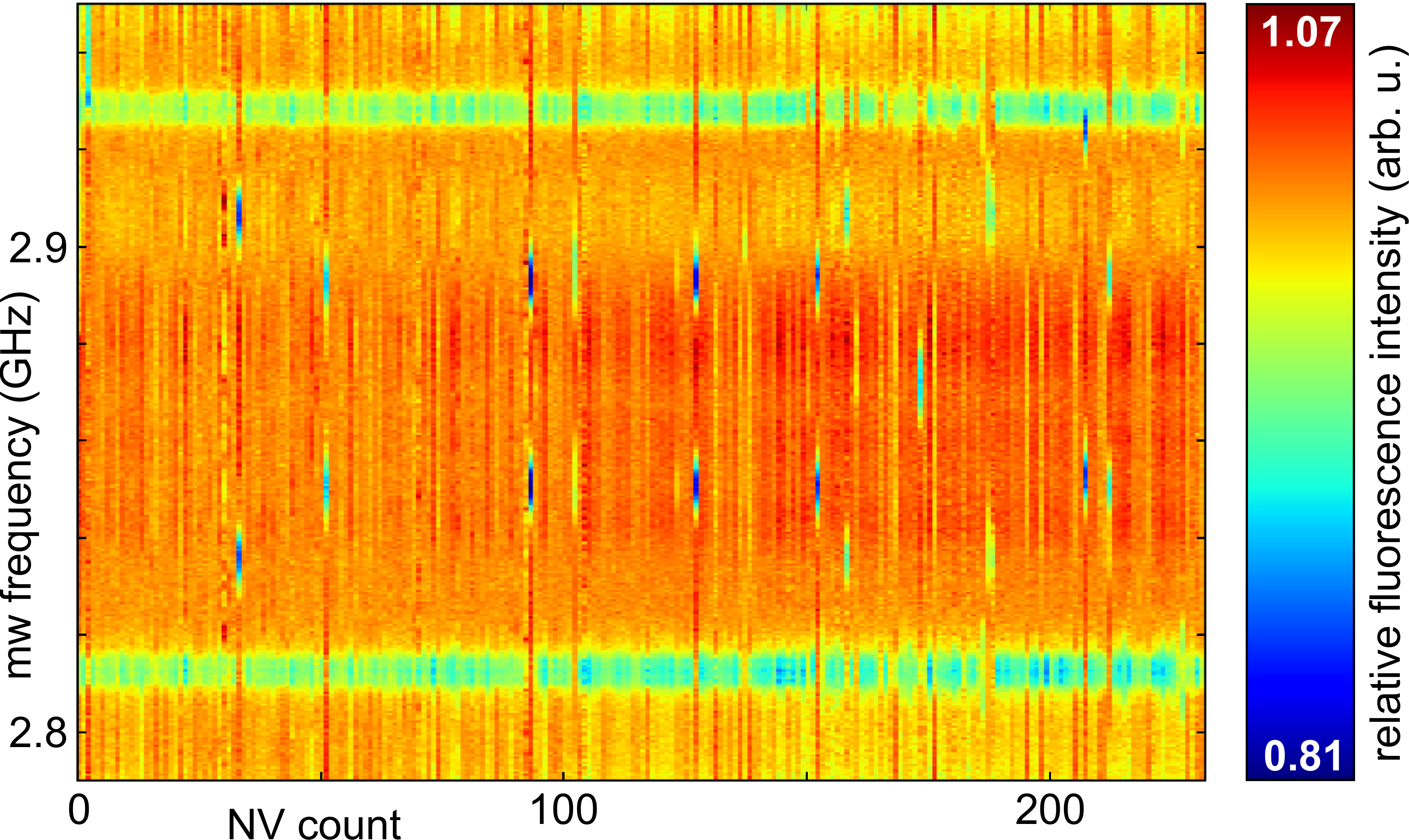}
  \caption{\label{fig:odmr}
    \textbf{ODMR reveals perfect NV alignment.}
		Horizontal stack of ODMR spectra (vertical) of 232 different NV centers (horizontal) in the CVD grown layer.
		The fluorescence intensity of the spectra is color coded.
		The dominant resonances (green horizontal lines)
		belong to NV centers that are aligned perpendicular to the diamond surface.
		Occasionally misaligned NV centers appear that have resonances which are spectrally within the two horizontal lines.
  }
\end{figure}

\paragraph{NV alignment via magnetic field application.}
Here we identify the NV alignment by taking ODMR spectra of randomly chosen NV centers in the optically characterized region of the CVD diamond layer.
To this end a magnetic field of $B\approx 4.1\,$mT is aligned roughly perpendicular to the diamond surface, i.e. along one of the $\left \langle 111 \right \rangle$ crystal axes.
According to eq.~\eqref{eq:H}, those NV centers which are aligned along the latter axis have a maximum Zeeman splitting of spin levels $m_S=-1$ and $+1$ of $2\gamma_{\mathrm{NV}} B \approx 115\,$MHz, whereas all other centers exhibit a considerably smaller splitting.
In fig.~\ref{fig:odmr} the spin transitions $m_S=0 \leftrightarrow \pm1$ are displayed for $232$ different NV centers in the CVD layer.
Almost all NV centers (218) show the maximum splitting (i.e. alignment perpendicular to the crystal surface).
Occasionally, some NVs (14) show a different alignment.
We conclude that at least $94\pm2\,\%$ of all NV centers within the CVD growth layer are aligned along the growth direction.

\begin{figure}[t]
	\includegraphics[width=1.0\columnwidth]{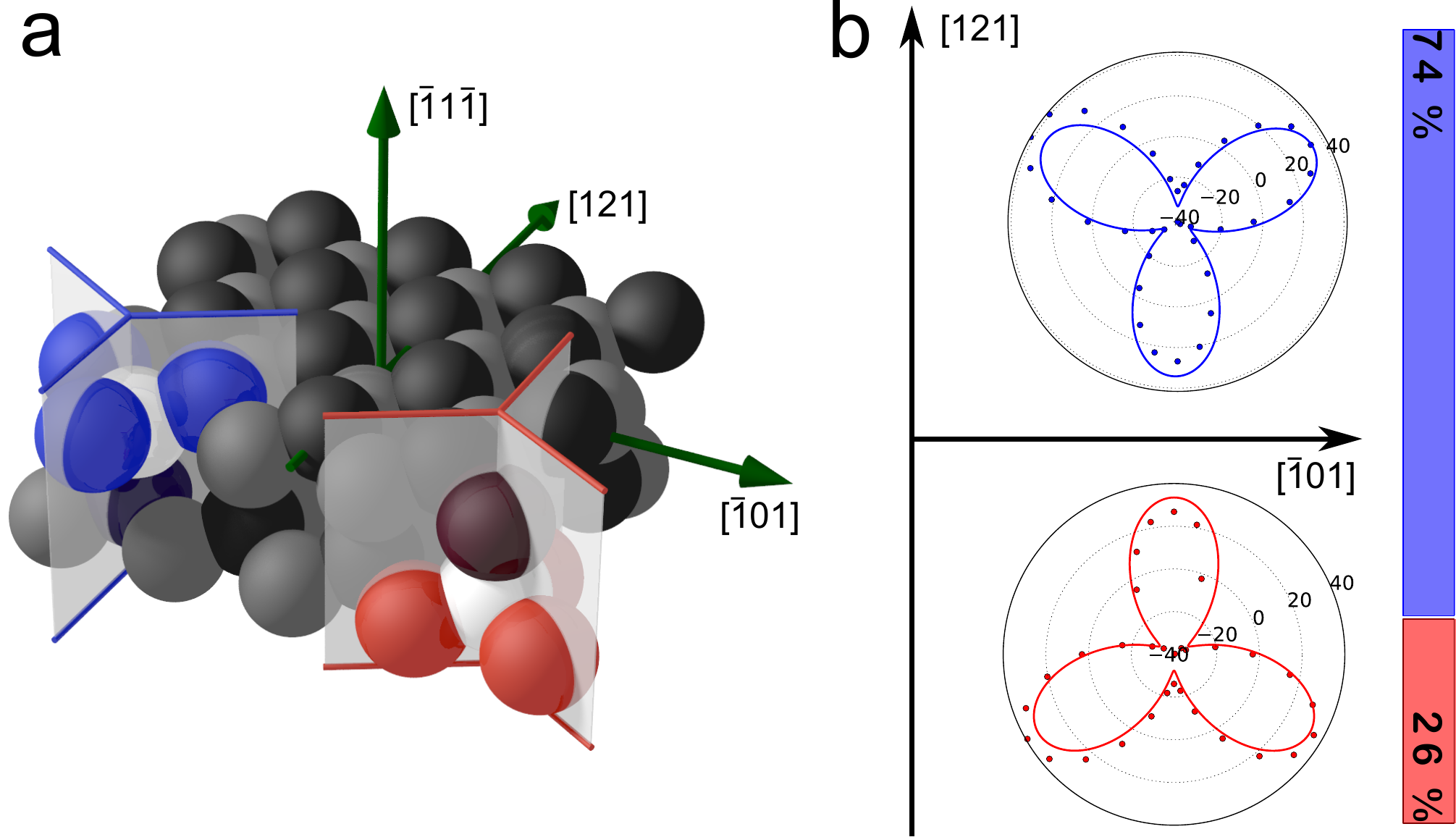}
  \caption{\label{fig:orientation}
    \textbf{Orientation of aligned NVs.}
    \textbf{a,} Crystal structure of the diamond sample.
		The two carbon sublattices are displayed in different shades of gray.
		Crystallographic directions are given where [$\overline{1}1\overline{1}$] (belongs to $\left\langle 111 \right\rangle$) is perpendicular to the sample surface.
		Two possible NV orientations with the perfect alignment are given (red with N up, blue with N down); vacancy is white and nitrogen is dark red or dark blue respectively.
		\textbf{b,} Polar plots of spin level shifts in kHz.
		The angle gives the direction of electric and magnetic fields. 
		The two distinct three-leaves correspond to the two orientations of the perfectly aligned NV centers.
		Of all investigated aligned NV centers $74\,\%$ were N down and $26\,\%$ N up.}
\end{figure}

\paragraph{NV orientation via combined electric and magnetic field application.}
We seek to also determine the orientation of the centers.
The simultaneous application of transverse parallel electric and magnetic fields reveals the threefold rotational symmetry of the NV spin Hamiltonian in eq.~\eqref{eq:H}.
The spin transition frequencies are
\begin{equation}
f_{\pm}=f_{\pm}\left(0\right)\mp k_{\perp}E_{\perp}\cos\left(3\varphi\right)
\label{eq:f}
\end{equation}
where $f_{\pm}\left(0\right)$ is only dependent on the magnitude of the transverse magnetic field,
$E_{\perp}$ the electric field strength and $\varphi$ the transverse orientation of the electric and magnetic fields \cite{dolde_detection_2013}. 
\\
\indent
At different $\varphi$, we perform ac-electrometry 
\cite{dolde_electric-field_2011} to determine $f_{\pm}$. 
The observed dependence of the lower transition frequency $f_{-}$ on $\varphi$ is displayed in fig.~\ref{fig:orientation} and corresponds to the trigonal structure of the NV center, where a lobe of the pattern is orientated in the direction from the vacancy to one of its nearest neighbor carbon atoms.
Two distinct orientations of the threefold pattern, which differ by a $\pi$-rotation, are found in the CVD layer of investigated NV centers.
\\
\indent
The diamond sample is a single diamond crystal and the crystallographic orientations of its faces have been characterized by X-ray crystallography, shown in fig.~\ref{fig:orientation}.
Two different orientations of NV centers with symmetry axes perpendicular to the surface exist.
Since the lobes of the threefold patterns correspond to the vacancy - nearest neighbor carbon directions of each NV center, the precise orientation of each NV center can be assigned.
\\
\indent
We have characterized 186 perpendicular NV centers in the CVD layer and have found a preferential orientation of $48:138$ for the nitrogen atom of the NV center pointing up vs. pointing down.
More precisely 74$\,\pm\,4\,$\%, have an orientation such that the nitrogen is located closer to the substrate.

\subsection{discussion}
We have shown experimentally that NV centers in as grown diamond layers produced by homoepitaxial MPCVD on \{111\} surfaces are not only perfectly aligned perpendicular to the surface, they are also mainly oriented such that the nitrogen atom is deeper inside the diamond than the vacancy.
\\
\indent
Very recently, the incorporation of substitutional nitrogen (N$_{\text{s}}$), NV and NVH defect centers during CVD diamond growth for typical surface orientations has been elucidated by quantum-chemical simulations \cite{atumi_atomistic_2013} in order to understand alignment and orientation of these as grown defects.
Regarding the formation of NV centers during CVD growth on (111) surfaces they conclude that N$_{\text{s}}$ in the topmost atomic double layer (see two uppermost layers in fig.~\ref{fig:orientation}a) is most stable, particularly in the displayed dark gray C positions.
Furthermore, N$_{\text{s}}$ is much more stable than a comparable NV center with the V in one of the neighboring light gray C positions.
This is consistent with the lack of misaligned NV centers in our CVD layer.
Furthermore, the findings in Ref.~\onlinecite{atumi_atomistic_2013} support our observed preferential orientation because N$_{\text{s}}$ atoms are supposed to grow in first during NV center formation.
Based on our findings, experimental growth conditions might be refined in order to improve the degree of preferential orientation, to increase the yield of such NV centers and the ratio of NV/N$_{\text{s}}$, which is expected to be tiny\cite{atumi_atomistic_2013}.
\\
\indent
Our presented technique and its potential for improvements will have direct impact on NV based quantum devices.
Particularly, in the envisioned hybrid quantum devices, where ensembles of NV centers are coupled to, for example, superconducting \cite{amsuss_cavity_2011,kubo_strong_2010}, optical\cite{jensen_cavity-enhanced_2014} and mechanical\cite{macquarrie_mechanical_2013,kepesidis_phonon_2013} resonators, a homogeneous NV ensemble is of great importance.
Where coupling to phonons is exploited, even the NV orientation will have a high influence on ensemble homogeneity.
Combined with recently developed techniques for the production of robust NV ensembles close to the diamond surface\cite{ohashi_negatively_2013}, the creation of aligned NV centers will greatly improve related sensors\cite{steinert_high_2010,steinert_magnetic_2013,le_sage_optical_2013}.
Finally, sensors based on NV ensembles in bulk diamond would benefit directly from our results because the established homogeneity will increase their signal to noise ratio by a factor of four as compared to no preferential NV alignment\cite{acosta_diamonds_2009}.

\begin{acknowledgments}
We thank I. Hartenbach for X-ray analysis and discussions as well as F. Jelezko for fruitful discussions.
We acknowledge financial support by DFG and JST (SFB-TR 21, FOR 1482), by the BMBF (CHIST-ERA), by EU (Diamant and SIQS) and ARC (DP120102232).
\end{acknowledgments}

\bibliography{RefsZotero}

\end{document}